\documentclass[11pt]{article} 

\usepackage{deauthor,times,graphicx}
\usepackage[T1]{fontenc}

\usepackage[colorlinks]{hyperref}
\usepackage{xcolor}
\usepackage{xspace}
\usepackage{subfig}
\usepackage{authblk}

\graphicspath{{xliu/}}

\makeatletter
\renewcommand\AB@affilsepx{, \protect\Affilfont}
\makeatother

\newcommand{\sysname}{\textsf{BeeTrace}\xspace}

\newif\ifsubmit
\submitfalse

\newcommand{\reviewer}[3]{
  \expandafter\newcommand\csname #1\endcsname[1]{
    \ifsubmit
      \ignorespaces
    \else
      \textcolor{#3}{[#2: ##1]}
    \fi
  }
}

\reviewer{evgenios}{Evgenios}{red}
\reviewer{dawn}{Dawn}{blue}
\reviewer{xiaoyuan}{Xiaoyuan}{orange}
\reviewer{nini}{Ni}{purple}

\begin{document}

\title{\sysname: A Unified Platform for Secure Contact Tracing\\ that Breaks Data Silos}

\author[1,2]{Xiaoyuan Liu}
\author[2]{Ni Trieu\thanks{nitrieu@berkeley.edu}}
\author[2]{Evgenios M. Kornaropoulos\thanks{evgenios@berkeley.edu}}
\author[2,3]{Dawn Song\thanks{dawnsong@berkeley.edu}}
\affil[1]{Shanghai Jiao Tong University}
\affil[2]{UC Berkeley}
\affil[3]{Oasis Labs}

\maketitle

\begin{abstract}
    Contact tracing is an important method to control the spread of an infectious disease such as COVID-19. However, existing contact tracing methods alone cannot provide sufficient coverage and do not successfully address privacy concerns of the participating entities. Current solutions do not utilize the huge volume of data stored in business databases and individual digital devices. This information is typically stored in data silos and cannot be used due to regulations in place. To successfully unlock the potential of contact tracing, we need to consider both \emph{data utilization from multiple sources} and the \emph{privacy} of the participating parties. To this end, we propose \sysname, a unified platform that breaks data silos and deploys state-of-the-art cryptographic protocols to guarantee privacy goals. 
\end{abstract}

\section{Introduction}

The lessons learnt from the recent events of the COVID-19 pandemic show the need to effectively track the chain of infection.
This task entails locating exposed individuals as well as identifying places that infected individuals have contaminated. 
The act of \emph{contact tracing via contact chain} can  significantly contribute in controlling the outbreak of an infectious disease. Unfortunately, traditional manual contact tracing approaches are not sufficient~\cite{Farrahi2014} against viruses that spread quickly via multiple transmission routes.  Digital contact tracing has the potential to be more effective by efficiently utilizing large amounts of data from multiple sources. In this work we focus on \emph{digital contact tracing}, as opposed to manual.

\textbf{Shortcomings of Current Approaches.} Most existing contact tracing applications focus on monitoring direct contacts between individuals. 
While this is a significant step forward, such an  approach does not consider the case of \emph{indirect contact} where individuals appear at the same location \emph{after} a contamination event. To address this phenomenon, we view contact tracing as a \emph{continuous monitoring procedure} that identifies both individuals who may have been exposed to the virus as well as high-risk places. 

The data sources used by existing contact tracing solutions are often limited, which in turn leads to coverage problems. Many contact tracing apps rely exclusively on GPS and/or Bluetooth for collecting contact information. Unfortunately, for the more challenging case of the indirect contact tracing, these sources are not enough. Additionally, these sources are not always reliable since GPS suffers from location precision, while Bluetooth techniques exchange tokens only between devices that are \emph{simultaneously at the same place} and as a result does not capture the case of indirect contact. The above state of affairs makes clear that we need \emph{additional sources} to improve coverage, e.g., data from business records. Interestingly, these needs have already been identified in manual contact tracing, where business data such as hotel and restaurant records are incorporated to discover potential contact with carriers. However, in the digital world, business records are stored in data silos, with records that are often in different formats, and are protected by strict legal regulations that forbid access to third-parties.

As a data collection and aggregation procedure, contact tracing needs to process \emph{sensitive data} of users' location and interactions. 
Therefore, privacy needs to be a key characteristic of a contact tracing platform. 
In order for the platform to notify exposed individuals it has to collect personally identifiable information (PII), along with the contact history and daily trajectory.  Exposure of this sensitive information in plaintext violates the privacy of users and exposes the participating parties to liability issues. To illustrate this point, an adversary that gets access to this information~\cite{corona-sniffer} can infer the daily habits of traced users as well as their social graph based on the contact history. 

\textbf{Our Contributions.} In this work we propose \sysname, a unified contact tracing platform that breaks data silos and achieves efficient data utilization while protecting privacy. We apply two key design elements: a \emph{unified data format with granularity adaptation} and a \emph{privacy-preserving distributed query framework}. 
Our contributions in this paper are the following: (1) We introduce a multi-source perspective to digital contact tracing systems and provide a way to track \emph{indirect contact} by specifying high-risk places as one of our tracing targets. (2) We put forth the setting of \emph{multi-source data collection} in contact tracing. In this setup we include business-side participation. (3) We list \emph{privacy requirements} and discuss information distribution options for the participating parties. (4) We incorporate all of the above insights in a unified contact tracing platform called \sysname. Our platform sets the vision for the standardization of the data format as well as the deployment of privacy-preserving distributed query processing. We provide use cases for a number of contact tracing scenarios and explain the corresponding cryptographic techniques, i.e., multi-party computation (MPC), and security protocols that meet our privacy requirements.

\section{Problem Definition}

    Contact tracing is a necessary control measure for highly infectious diseases~\cite{Fraser2004} that typically present a high pre-symptomatic ratio and a high basic reproduction number ($R_0$), i.e., the average number of infections caused by a single carrier. In the recent case of COVID-19, the source of transmission may come from droplets, aerosol, contamination of surfaces, and fecal-oral contamination~\cite{Ferretti2020}. As a result, once an infectious carrier contaminates a place, the risk of infection remains for a prolonged period of time.

\begin{figure}
\centering
\includegraphics[width=0.77\textwidth, bb= 0 0 1829 1005]{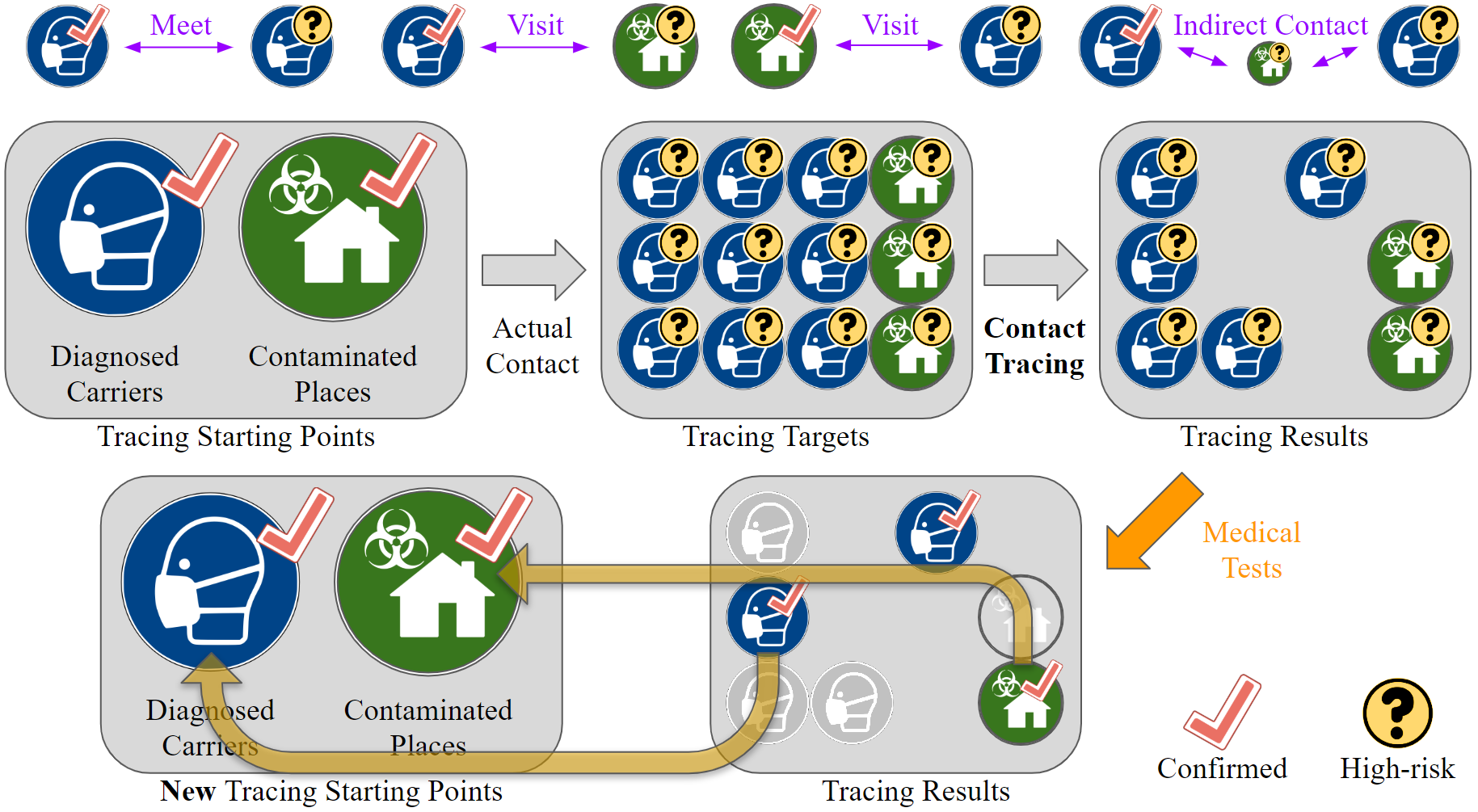}
\caption{An illustration of the Contact Tracing process and its key terms. The top row presents potential direct and indirect contact scenarios. 
A contact tracing system needs to start from the confirmed cases defined as \emph{tracing starting points} and  compile the \emph{tracing targets} that get a medical test to produce the \emph{tracing results}. As a last step, newly diagnosed carriers and contaminated places are added to the tracing starting points.}
\label{problemdef}
\end{figure}

\textbf{Terminology.} In this work we consider both \emph{static places}, such as hotels and restaurants, as well as \emph{dynamic places}, such as buses and taxis. 
We use \emph{diagnosed carriers} to refer to individuals who have positive medical test results and have not recovered yet. 
Asymptomatic, pre-symptomatic or mildly-symptomatic infectious individuals are often unaware of their infection before taking a medical test. We call these individuals \emph{undiscovered carriers}. Since both diagnosed carriers and undiscovered carriers are infectious, we consider all the places they have visited as \emph{high-risk places}. 
We describe the process of contact tracing  using the following terminology, see
Figure \ref{problemdef} for an illustration:

\begin{itemize}
    \item \textbf{Tracing Starting Points.} This is a list of verified carriers as well as places for which a confirmed carrier visited. All entries of this list are accompanied by an official record that proves the authenticity of their contamination claim, e.g., for the case of a carrier a medical test result and for a case of a contaminated place a business transaction initiated by a confirmed carrier. 
    \item \textbf{Tracing Targets.} This is a list of individuals and places that interacted with an entity from the list of tracing starting points. This list includes the following entities:
    \begin{itemize}
        \item Individuals that have been within certain distance from a diagnosed carrier from the list of tracing starting points.
        \item Individuals that have visited a contaminated place from the list of tracing starting points.
        \item Places that have been visited by a diagnosed carrier from the list of tracing starting points.
    \end{itemize}
    \item \textbf{Contact Tracing.} This functionality takes as an input the tracing starting points, and outputs the the tracing targets. We use the term \emph{tracing result} to refer to the output of the contact tracing functionality. A successful contact tracing should have a tracing result that matches precisely the true tracing targets.
\end{itemize}

A \emph{contact tracing system} is an automated system that uses data from multiple sources to conduct contact tracing. Our vision for large-scale contact tracing involves (1) a \emph{unified approach} of processing data from \emph{multiple sources}, (2) and a focus on the  \emph{privacy requirements} for all participating individuals and businesses.

\subsection{Contact Tracing from Multiple Sources: Necessity of Breaking Data Silos}
\label{bds}

A unified contact tracing system needs to break data silos and use multiple data sources to achieve high coverage. Figure~\ref{fig:example} illustrates with a concrete example the necessity of breaking data silos. We note here that this is the running example that we use across our work and it is used for motivating our vision for \sysname.

\begin{figure*}[t!]
	\centering
	\framebox{
	    \noindent\fcolorbox{black}{gray!20}{
    		\begin{minipage}{0.95\linewidth}
    		    One sunny morning, Alice decides to visit her friend in a distant town. After taking a commercial flight to her friend's town, she calls a taxi using her mobile phone and goes to a famous restaurant for lunch with friends. Alice and her friend then go for a walk at a nearby park before parting ways. She takes the subway from the park to the airport but is informed that her flight was delayed. She changes her plans and decides to stay overnight at a nearby hotel. The next day,
    		    \begin{itemize}
    		        \item Alice receives a notification on her phone to self-isolate and seek a medical test.
    		        \item (Or) the hotel staff is also aware that Alice is at risk and calls to adviser her to self-isolate before taking a medical test.
    		    \end{itemize}
    		\end{minipage}
    	}
	}
	\caption{Real-world example of contact tracing based on Alice's outing.}
	\label{fig:example}	
\end{figure*}

In the following we list potential events from which a newly confirmed carrier, whom we call Alice, was infected either through direct or indirect contact by a known carrier, whom we call Bob.
Then, given the events that took place during Alice's outing, see Figure~\ref{fig:example}, one of the following events took place: (1) Bob took the same taxi earlier that day, (2) Bob had lunch at that restaurant, (3) Bob walked at a very close distance to Alice at the park, (4) Bob was at the same subway car as Alice, (5) Bob stayed at the same hotel room.

Without records from multiple sources, e.g., transactions from the taxi company and the restaurant, it is highly unlikely that a contact tracing platform can identify which of the above events took place. As a result, users such as  Alice would not have been notified and would probably spread the infection even further. 
To address the challenges of multi-source contact tracing we use \emph{existing (traditional) data sources} and we create \emph{new data sources} tailored to task of contact tracing. 

\textbf{Existing Data Sources.} 
Notice that existing databases from businesses that Alice visited already record her contact history with both individuals and places. We envision a platform, namely \sysname, that can \emph{facilitate a secure computation over multiple service providers} such as ride-sharing companies, subway companies, restaurant businesses, and hotels, in order to perform accurate, scalable, and privacy-preserving contact tracing.

\textbf{New Data Sources.} As for new data sources that are focused on contact tracing data, one  can either incorporate data from newly developed applications such as~\cite{apple-google-ppct} or develop an independent contact tracing application.  
Most new applications use either Bluetooth or sensors such as mobile-phone camera for QR code scanning. 
In order to cover high-risk place tracing using Bluetooth, one potential approach is to install \emph{Bluetooth beacons} in taxis, subways, and restaurants. These beacons transmit random identifiers that can be later found on a carrier's device. 
Such an approach is compatible with solutions like Google/Apple exposure notification systems~\cite{apple-google-ppct}. This solution has a low implementation overhead if one is to follow the same format (Rolling Proximity Identifier) as the protocol of a deployed notification mechanism. Another approach is to put QR-codes in public places that can be scanned with a mobile phone camera and accurately locate the time and the location of a participant.  
This technique is successfully applied in China~\cite{China-qrcode}. Both approaches can contribute to indirect contact tracing.

\textbf{Proposed Approach.} Interestingly, one can decouple the step of \emph{collecting data} from the step of \emph{aggregating data}. 
We only consider the case where the data that is collected by each participant is not transmitted to a centralized authority but rather stays in a thin client that is part of our platform. 
Specifically, our approach can integrate data from existing contact tracing apps into our aggregation platform for comprehensive coverage. 
Our platform can also integrate data from individual users' mobile phones and can be extended to consider other data sources including data sources collected by businesses. To achieve this functionality we propose the conversion of the data generated by third-party apps and we use them as a contact tracing input to our platform. A challenge that we had to overcome is that both the \emph{precision} and the \emph{format} of third-party generated data varies.
A key-design of our platform is \emph{a unified data format with granularity adaptation}, which is discussed in Section \ref{dataformat}.

\subsection{Privacy Requirements and Information Distribution}
\label{options}
Privacy is a central issue in contact tracing platforms especially since the most essential data is the most sensitive data, e.g., geolocation of users, health records of carriers, business transactions. 
Societies have already witnessed the consequences of exposing sensitive information of participating individuals, e.g.,~\cite{south-korea-privacy}. 

Designers of contact tracing platforms need to consider the following three questions: (1) Where does the input data go? (2) Who performs the contact tracing computation on the input data? (3) Who is the recipient of the output of the contact tracing computation?
A first attempt is to put all data into a single central server but such an approach introduces a single point of failure and  \emph{exposes sensitive data} to whoever has access to the server. 
Jumping ahead, our platform is based on cryptographic techniques and scalable secure protocols~\cite{pet18, Kales2019, SOda19} that \emph{do not reveal any information to the participating parties other than the result of the computation}.

\textbf{Distributing the Result of Contact Tracing.} 
The options for distributing the contact tracing results vary depending on which party needs to be notified. 
We consider three major parties: the \emph{medical authorities (MA)}, the \emph{businesses}, and the \emph{users}. There are two main goals for the \sysname platform:

\begin{itemize}
    \item \textbf{Broadcast of High-Risk Places}. The MA publishes heat maps that depict the high-risk places. 
    This list can help individuals to independently evaluate the risk of infection and self-report when in need for a medical test. Sharing this data with the community contributes in data-informed decisions about daily travel routes. Thus, users have the option to avoid high-risk areas and means of transportation. 
    \item \textbf{Targeted Notifications}. The \sysname\ platform provides continuous updates for every participating user. It takes MA-reported diagnosis as input and distributes the tracing results to the relevant targets. Users that are part of the tracing result are notified by \sysname\ and can proceed with taking a medical test. We emphasize that the notifications are sent directly to the relevant party without exposing to MA the identity of the user at risk, which protects the privacy of \sysname\ users.
\end{itemize}

We propose a decentralized approach, which is a main key design of our \sysname\ platform: \emph{a privacy-preserving} and \emph{distributed query framework}, which is discussed in more detail in Section \ref{ppframework}. 
\emph{Our privacy-preserving design focuses on introducing no extra privacy leakage, while improving the efficiency of the contact tracing.} 
We enable this functionality using secure multi-party computation (MPC) techniques which are shown to be practical in other deployed real-world systems.

\textbf{Privacy Requirements. }In addition to the information distribution requirements, one needs to consider privacy requirements such as, (1) \emph{Anonymity requirement}: Avoid both explicit and implicit PII exposure. It should prevent linkage attacks and intersection attacks. 
(2) \emph{Informed consent requirement}: Participants have the right to know what information is provided and how the information is processed. (3) \emph{Confidentiality requirement}: All provided information should be protected against unauthorized access. 
(4) \emph{Limited-time storage requirement}: All locally-stored information must not be kept longer than needed. (5) \emph{Right to be forgotten requirement}: Participating individuals and businesses should be able to retract all their information generated through direct and indirect contact.

\section{\sysname: Platform Overview}
\label{poverview}

\begin{figure}
\centering
\includegraphics[width=0.8\textwidth, bb= 0 0 1839 669]{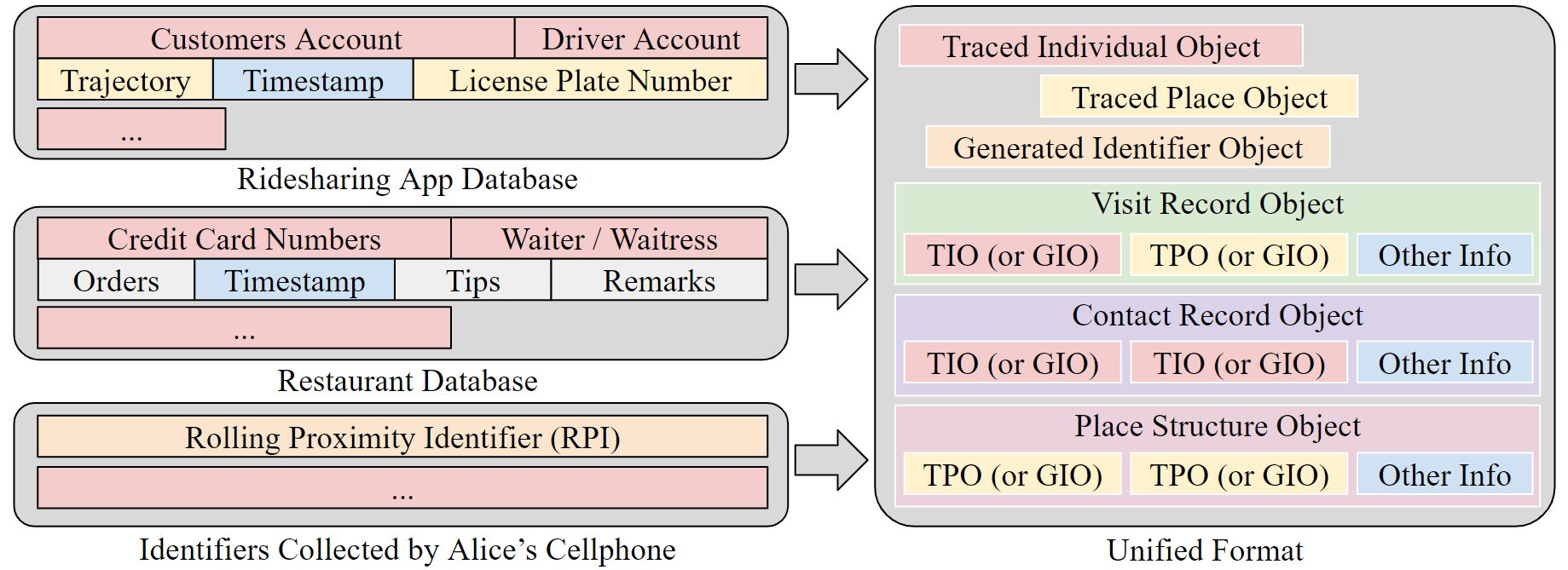}
\caption{Format conversion between business data and our proposed unified format. On the left, we have records from businesses and third-party contact tracing apps with different format. All contact tracing related information from these records can be converted and stored locally in our unified format.}
\label{busidata}
\end{figure}

The \sysname platform follows two new key components. 
The first component is a \emph{unified data format with granularity adaptation} and introduces a new unified format that is comprised of a series of objects, see Figure~\ref{busidata}.
We address the challenges of multi-source contact tracing presented in Section~\ref{bds} by proposing this unified format which captures all the useful attributes that are generated by existing and new data sources. 
The second component is a \emph{privacy-preserving distributed query framework} that allows the execution of queries over the objects of the unified data format. 
We note here that the result of each query is distributed only to relevant parties as described in Section~\ref{options}. 
To simplify the exposition in this section, we refer to unified data format and secure protocols without explaining the terms in detail. 
We note that a more thorough presentation of these two terms is provided in Section \ref{dataformat} and Section \ref{ppframework}.

In the following we present the workflow of the proposed platform \sysname. 
Our platform consists of two phases: the \emph{setup phase} and the \emph{operating phase}. 
The description of both phases is presented from the perspective of the three participating parties, the medical authorities (MA), the businesses, and the users.

\subsection{Setup Phase}

The first step that a party takes towards joining the \sysname platform is to download and install the \emph{client} application of the platform according to the party's type. In particular, there are three types of client applications---MA-side client application, business-side client application, and user-side application.
We assume that all parties operate under the semi-honest security model and they use their real data as an input for the secure multiparty computation protocols.

\textbf{Medical Authorities. }In the setup phase, the medical authority (MA) simply installs the client of the platform. 
As a next step, the newly installed client converts all the information stored locally in the database of MA into the proposed unified data format. 
We note that the database of the MA at setup time contains all the verified infections that were collected \emph{before} the MA joined the platform.

\textbf{Businesses. }The businesses that voluntarily participate in \sysname have to first install the business-side client application of the platform. For simplicity, we assume that there are no malicious actors that impersonate a business; we note that one can deal with such a scenario with known mechanisms such as Certificate Authorities.
As a next step, the business converts locally each transaction that is stored in the local database into a unified data format entry. 
For example a hotel converts the stay of each visitor into a new unified data format entry. 
The rationale behind this step is that in future operating phases that take place \emph{after the setup}, the hotel has already converted its information to the appropriate format and can participate in the privacy-preserving distributed query execution.  
We emphasize that the unified data format entries never leave the premises of the business and only participate in secure computation protocols that do not reveal the input of each party.

\textbf{Users. }The users that choose to participate in the platform have to first download user-side client application for their mobile device. As a next step the application converts the data collected from the device's sensors to the unified data format. 
For example, this data includes GPS traces and Bluetooth tokens. 
Similar to before, this conversion is performed so that the client can participate in the privacy-preserving distributed query execution and the plaintext unified data format entries never leave the premises of each user.

\subsection{Operating Phase}

In the operating phase, the participating parties collectively and continuously engage in privacy-preserving distributed query executions. 
The setup phase already converted the previously-generated data of each party to the unified data format. 
Every future data generation is converted to a unified data format entry automatically through the local client application. 
At a high-level, every discovery of an infection as an expansion of tracing starting points triggers the distributed execution of a query across all participants to finally generate the notifications.
Our approach creates a \emph{closed loop for continuous multi-source monitoring of the chain of infection}.

\textbf{Medical Authorities. }In the operating phase, the MA continuously receives reports about diagnosed carriers and contaminated places in our unified format.
For each diagnosed carrier, the MA only learns a pseudonym or some tokens , i.e., an opaque ID or  Rolling Proximity Identifiers.
  These unified data format entries in received reports are added to the list of tracing starting points. 
In turn the knowledge of the newly infected party triggers the execution of a query so as to discover the participating parties that came in direct and indirect contact with the carrier, i.e., discovery of tracing targets. 
Another responsibility of the MA is to collectively compute a high-risk heat map of the known infections so as to inform the community. 
The computation of this heat map takes place using cryptographic techniques that only output the aggregate information without revealing the identity of the infected individuals. 
The resulting map of this process is broadcast to all participating parties in the platform.

\textbf{Businesses. }The participating businesses use their locally stored data in unified format to contribute in the distributed privacy-preserving computation that is triggered by the contact tracing cycle. 
If a verified carrier completed a transaction with a participating business, then the platform notifies the manager of the place that the business is part of the list of tracing results. 
In turn the business shares the unified data format entries of the rest of the clients that performed a transaction within a certain time-frame. The affected users are now also part of the tracing results and, after further medical tests, the contact tracing continues. 
The business follows the direction of the MA to decontaminate its premises and removes itself from the tracing starting points list after the prescribed period of time.
 With the consent of the user, the business may also actively notify the user about their risk using communication methods already existing in their businesses.

\textbf{Users. }The participating users contribute in the distributed privacy-preserving computation with their data that capture, among other things, the GPS trajectories, collected Bluetooth tokens, visited businesses, precise QR-code generated geolocation etc.
If a user is added to the tracing results via a confirmed direct or indirect contact then the platform directs the user to the closest medical test center. 
If the user is indeed a newly infected case, then the platform triggers another distributed query that updates the list of tracing starting points.

\section{Unified Data Format with Granularity Adaptation}
\label{dataformat}

In this section we give a detailed road-map of our vision about a unified data format. 
The first step towards breaking data silos is the creation of a \emph{unified representation} for contact tracing data. We need to convert data from different sources to a universal format before processing it. 
Another challenge that we need to address is that spatial and temporal data is often in different granularity. 
Therefore, we need to create the corresponding abstraction to capture all generated data.

\textbf{An Approach for Securely Unifying Data.} For business purposes, the data stored locally at each business party typically contains information related to individuals, e.g., full name, username, registration email, credit card number etc.
Our privacy requirements do not permit the uniquely identifiable information to leave the premise of the participating business/user. 
 We introduce the notion of an \emph{opaque ID} to represent each piece of  sensitive information. 
This identifier does not reveal anything about the true identity of the entity it refers to. 
We note that opaque IDs can be generated by one-way hash function.

To demonstrate the \emph{power of the proposed unified data format} we list a series of data objects that are relevant to typical contact tracing scenarios. 
We propose intuitive object descriptions to facilitate an easy conversion between original data and our proposed format. 
With the term \emph{domain object} we denote a record associated with an opaque ID which in turn refers to either a user or a place.  
With the term \emph{relation object} we denote a record that describes the relation between a pair of objects, e.g., the relation between a subway car and a subway train.

\textbf{Domain Objects.} We identify that the following domain objects are typically used in contact tracing:
        \begin{itemize}
            \item \textbf{Traced Individual Object.} It uses a set of key-value pairs of strings to describe the real-world information of an individual which will be used later in matching during contact tracing. It also contains a field for infection status and a field for risk evaluation status. The descriptive information can contain names, email addresses, phone numbers, or just usernames in apps depending on what information has already been collected for business purposes. All the values are secured by opaque ID to further protect user privacy. We note that a traced individual object may contain many fields, this design choice allows matching based on data that comes from different sources. 
            
            \item \textbf{Traced Place Object. }These are the objects for traced places. 
            They describe a static or dynamic place, and each object contains a field for contamination status. 
            The description can be tailored to cover different location granularities, e.g., room of a hotel, floor of a hotel, hotel building, street of the hotel. We use place structure relation objects, detailed in the next paragraph, to explain the \emph{belonging relation} between traced places.

            \item \textbf{Generated Identifier Object. }This object contains tokens generated by BLE or other contact tracing apps. These tokens can be associate to either individuals or places. Generated identifier objects are \emph{ephemeral} and often have an expiration date different from the first two types of data  objects. They also contain a field for risk evaluation. To illustrate its generation, when our client collects Rolling Proximity Identifiers from the Google/Apple solution we convert them to our unified format and use them for querying.

        \end{itemize}
 \textbf{Relation Objects. }These objects tie two distinct parties so as to keep track of either \emph{direct contact} or \emph{indirect contact}. We note that depending on the circumstances it is plausible that only one of the two parties generated a relation object. To overcome this asymmetry, we explain in the next section how to perform distributed queries to discover the relation objects that users/places unknowingly participate in. 
 We identify that the following relation objects are typically used in contact tracing scenarios:
        \begin{itemize}
            \item \textbf{Visit Record Object}: This is a record that describes the event where an individual visited a place at a specific time (with varying precision granularity). It usually contains one traced individual object (or generated identifier object) and one traced place object (or generated identifier object). Visit record objects are created when businesses have an entry in their database describing this visit, e.g., a customer checking in at a hotel or a customer getting on a car. They are also created periodically by user-side client application using GPS sensors. 
            \item \textbf{Contact Record Object}: This is a record that describes the event of two individuals being within a certain distance at a specific time (with varying precision granularity). It usually contains two traced individual objects (or generated identifier objects). For example, the ride-sharing companies create contact record objects for the driver and the customer, and the restaurants create them for customers who seat close. The user-side client application also collects generated identifier objects from existing contact tracing apps, e.g., ones that contain the Rolling Proximity Identifiers, and create contact record objects for them with the timestamp attached.

            \item \textbf{Place-Structure Object}: This record describes the relation between two traced places. For example, we want the system to understand a room in a hotel (which is a traced place) is in that hotel (which is another traced place). This comes handy, when we have a record of a carrier visiting the hotel without knowing which room (s)he stayed at. With place structure object, we can automatically infer that all traced places in this hotel are at high risk. Place structure objects form a tree-like structure for all traced places. E.g. room 301 is on the third floor of a hotel, which belongs to a certain hotel,  which is part of a street block.
        \end{itemize}

As discussed in Section~\ref{poverview}, each party will first download and install the client application of the platform. After the installation, the MA import all existing data about diagnosed carriers and contaminated places, and the MA-side client application will create the corresponding domain objects which are labeled infected. The business creates place-structure objects using the user interface of business-side client application and also import their contact tracing related data or connect its client application to its databases for continuous object creation. The user inputs their personal information and the user-side client application will generate traced individual objects secured by opaque ID for later matching. The user also turns on the sensors and grant app data access permissions so that user-side client application will start to collect and save domain objects and relation objects. It use GPS and QR code scanning to collect visit record objects, and use Bluetooth and other contact tracing apps to collect contact record objects, e.g., the Rolling Proximity Identifiers are first converted to generate identifier objects and are used for matching later.

\begin{figure}
\centering
\includegraphics[width=0.8\textwidth, bb= 0 0 1280 721]{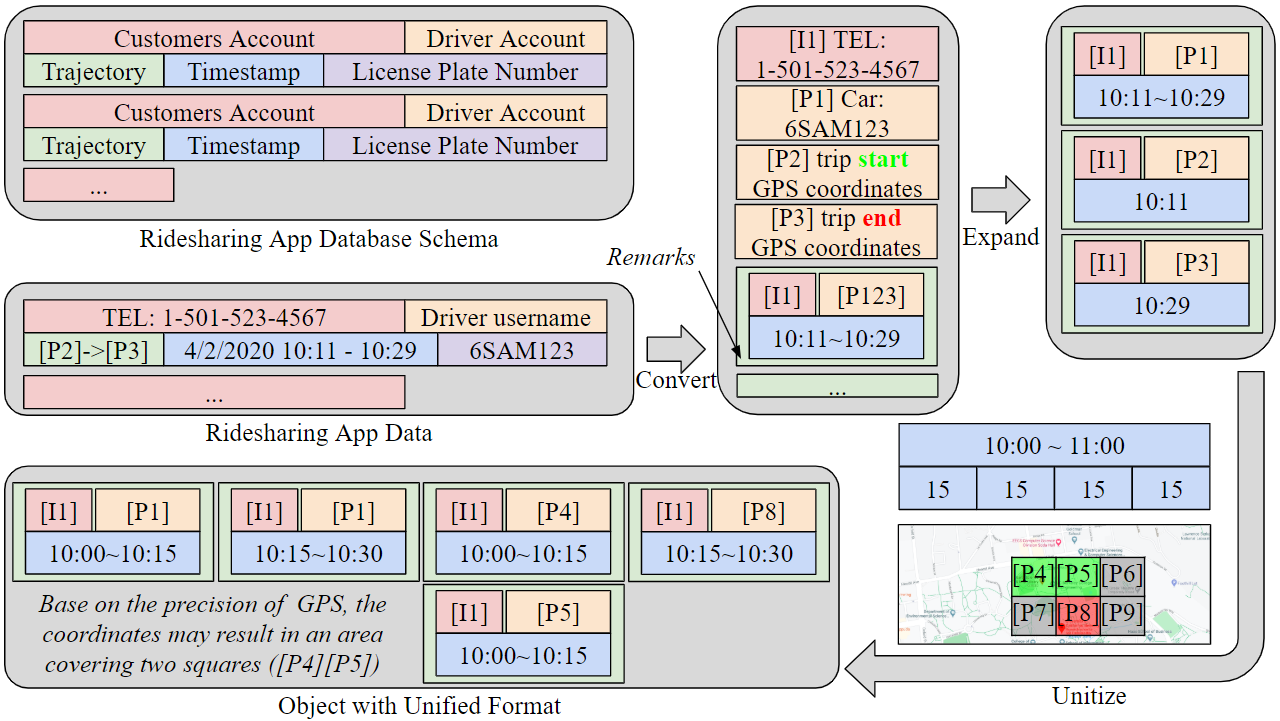}
\caption{Illustration of expansion and unitization. The data from ride-sharing app is first converted into objects in our unified format. In the expansion step, the \emph{visit record object} is expanded to three visit record objects that cover the start location, the end location, and the dynamic place of the car. Then in unitization step, we split one hour into four segments and generate the visit record objects accordingly.}
\label{procstep}
\end{figure}

We note that in each object description the \emph{granularity} of the place and time may vary. 
We use two techniques to convert objects to the same granularity: \emph{expansion} and \emph{unitization}. Figure \ref{procstep} presents a concrete example of these two steps using our motivating scenario with Alice. 

In the \emph{expansion step}, we expand an object by splitting the object into the smallest granularity. 
Given a traced place object, we analyze its tree structure implied by the  place-structure objects to locate the smallest granularity.
E.g., for GPS coordinates, we create a tree structure by superimposing grid cells (see the map in the lower right corner of Figure \ref{procstep}).

In the \emph{unitization step} we process the continuous data entries, e.g., coordinates, timestamps. 
The continuous data may have a large domain; to limit the output space we split the domain into multiple segments and map the data accordingly. 
The number of segments is a tunable parameter of the platform. 
For example, in the figure we use fifteen minutes as the hour unit and split 24 hours into 96 segments with the same length. 
By processing the data under the same basic unit, we can seamlessly execute queries under a consistent domain of values.

\section{Privacy-preserving Distributed Query Framework}
\label{ppframework}

In this section, we introduce the proposed framework of platform \sysname for \emph{privacy-preserving distributed data queries}. We first formalize the application scenarios for both targeted notification and high-risk place broadcast, then discuss security protocols that enable these applications. 

In this work, we consider a \emph{semi-honest} security setting where the adversary is assumed to follow the protocol, but attempts to infer additional information from the execution of the protocol. We assume there are multiple businesses and multiple users in our \sysname\ but only a single MA. 

The Generated Identifier Objects, see Section~\ref{dataformat}, may refer to either traced individuals or traced places. 
To illustrate this ambiguity, notice that a Bluetooth beacon installed at a place may use the same format as the Rolling Proximity Identifier which is tied to individual and, consequently, we may not be able to distinguish these two scenarios. 
As a solution, when exchanging data related to either individuals or places, we assume Generated Identifier Objects to be both and always include them.

\subsection{Targeted Notifications: A  Privacy-Preserving Approach}
There are two targeted notifications, a notification towards a business, and  a notification towards a user. The first case takes place when the MA assists businesses in resolving  whether they are a high-risk place. The second case takes place when the MA and businesses collectively use their data to assist a user in resolving whether they are in the tracing results. Concretely,

\begin{itemize}
    \item[(1)] \textbf{Notification Towards a Business.} The MA has domain objects that describe individuals and places in the tracing starting points.  A business has a list of domain objects and wants to find out if any of its domain objects appear in the tracing starting points. 
    
    \item[(2)] \textbf{Notification Towards a User.} A user has a list of domain objects related to (or generated  by) himself or the individuals (s)he met, e.g., the Rolling Proximity Identifiers from google/apple solution, and the places (s)he visited. The user wants to know if any of these objects matches the objects with infection risk or diagnosis from MA or businesses. 
\end{itemize}

\sysname performs targeted notification while maintaining the privacy of each participant. 
To solve this problem in a privacy-preserving manner, we deploy an efficient cryptographic primitive called \emph{ private set intersection (PSI)}. 
The setup of PSI is the following: there are two parties, the sender and the receiver, and each party holds a set of elements. 
The PSI functionality allows the receiver to learn the intersection set and nothing else. 
Some real-world applications have already adopted the PSI primitive and achieve practical performance~\cite{Kales2019}. 
Using the PSI functionality, our \sysname\ can provide solutions for (1) and (2). 

For (1), the MA and the business execute a PSI instance where the desired output of this protocol is the \emph{tracing result}, i.e., the intersection between sets.
Each party provides as an input a collection of locally-stored domain objects (e.g, traced place object). At the end of the PSI, the business who acts as PSI receiver learns the tracing result and nothing else, which protects the raw tracing starting points. On the other hand, the PSI gives nothing to the PSI sender MA which protects the business's input. Having the tracing result, 
the business can notify their customers about the newly discovered risk, e.g., the hotel in our example scenario, or just mark the status of their locally stored objects and wait for queries from users.

Similarly, for (2), the user executes a PSI instance with the MA as well as all businesses that (s)he has visited. With the list of domain objects as input from each party, PSI allows the user to learn if any of its objects appears in the object lists of MA or the businesses. Because of the cryptographic guarantees of the underlying PSI protocol, no information is revealed except the PSI output. 
If the intersection is non-empty then the user may potentially be infected. We emphasize that our platform also discover the risk that can be detected by different existing contact tracing apps. Because the generated identifier objects are also domain objects, and by matching them, e.g., matching the Rolling Proximity Identifiers, the user-side client application can deduce if the user is at risk.

As an alternative if we only want a business to check its contamination status \emph{without revealing which customers was infected}, we can use a variation of PSI called PSI-CA~\cite{Ni2020}. 
From the output of this protocol the business only learns the number of infected users that visited its premises without identifying their opaque ID. 
An important technical detail of our approach is that we need to choose records with reasonable timestamp before issuing a query in order to reduce the false-positive alerts to customers, which is especially important when the query contains a large number of generated identifier objects describing ephemeral tokens like Rolling Proximity Identifiers.

When using PSI and PSI-CA to handle the generated identifier objects, the method is different from the google/apple approach but the ephemeral tokens, like Rolling Proximity Identifiers, are still used in a privacy-preserving way. The user-side application deduces the risk of individual infection by executing a PSI instance with the MA to check if the user has a matched token, which indicates whether the user has direct contact with someone in the tracing starting points. With tokens generated by Bluetooth beacons, this can be generalized to tracing for places similarly to cover indirect contacts.

\begin{figure*}[t!]
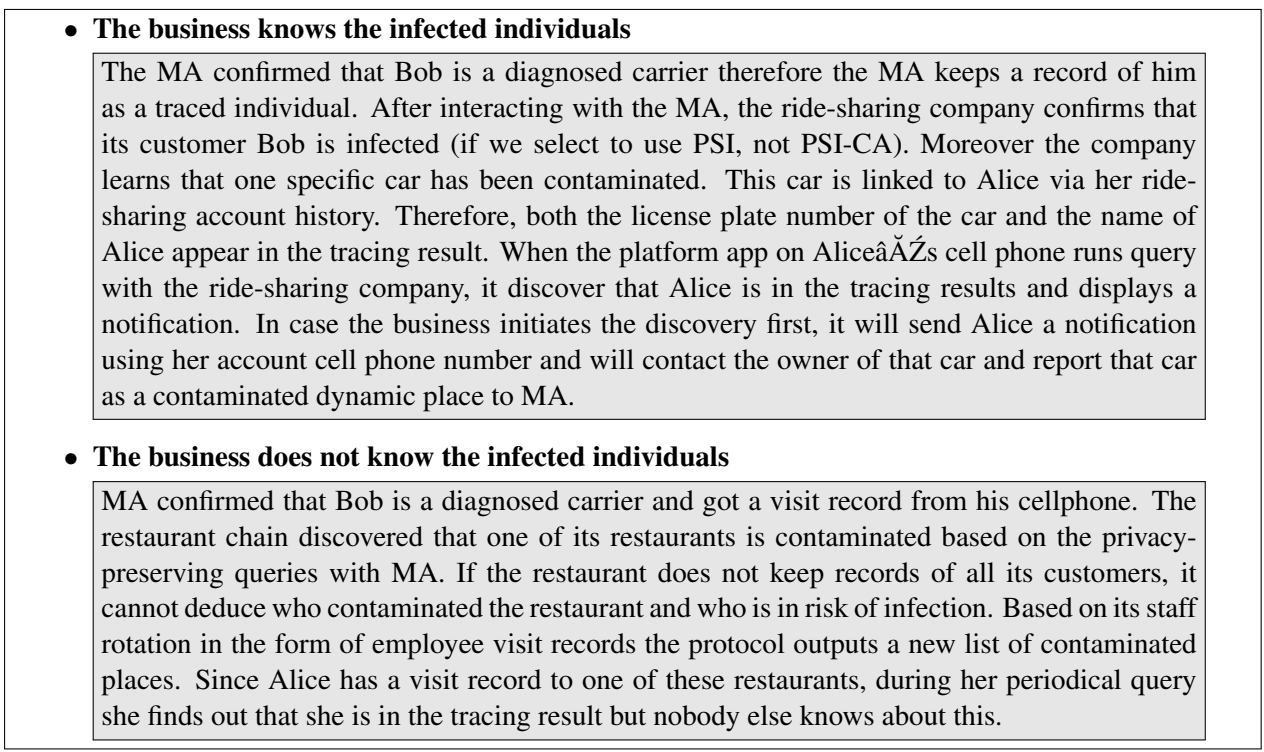

	\centering
	\framebox{
    	\begin{minipage}{0.95\linewidth}
    	    \begin{itemize}
    	        \item \textbf{The business knows the infected individuals}

\noindent\fcolorbox{black}{gray!20}{\begin{minipage}{0.95\linewidth}
    The MA confirmed that Bob is a diagnosed carrier therefore the MA keeps a record of him as a traced individual. After interacting with the MA, the ride-sharing company confirms that its customer Bob is infected (if we select to use PSI, not PSI-CA). Moreover the company learns that one specific car has been contaminated. This car is linked to Alice via her ride-sharing account history. 
    Therefore, both the license plate number of the car and the name of Alice appear in the tracing result. When the platform app on Alice’s cell phone runs query with the ride-sharing company, it discover that Alice is in the tracing results and displays a notification. In case the business initiates the discovery first, it will send Alice a notification using her account cell phone number and will contact the owner of that car and report that car as a contaminated dynamic place to MA.
\end{minipage}}
    	        \item \textbf{The business does not know the infected individuals}

\noindent\fcolorbox{black}{gray!20}{\begin{minipage}{0.95\linewidth}
MA confirmed that Bob is a diagnosed carrier and got a visit record from his cellphone. The restaurant chain discovered that one of its restaurants is contaminated based on the privacy-preserving queries with MA. If the restaurant does not keep records of all its customers, it cannot deduce who contaminated the restaurant and who is in risk of infection. Based on its staff rotation in the form of employee visit records the protocol outputs a new list of contaminated places. Since Alice has a visit record to one of these restaurants, during her periodical query she finds out that she is in the tracing result but nobody else knows about this.
\end{minipage}}
    	    \end{itemize}
        \end{minipage}
	}
	\caption{Example for the privacy-preserving distributed query framework in \sysname\ }
	\label{diffviz}	
\end{figure*}

Recall that the proposed platform \sysname\ aims to protect the privacy of participants while maintaining the contact tracing functionality. 
We illustrate this goal with two examples of distributed queries in Figure \ref{diffviz}.

\subsection{Private Discovery of High-Risk Place}

To enable the discovery of high-risk places,  the MA wants to learn which area the diagnosed carriers have visited while the  carriers do not want to expose their own trajectories. In fact, the MA only needs the aggregated result, which does not expose information of an individual user. 
We first formally define the problem and then discuss two solutions that give different efficiency and privacy trade-offs.

\begin{itemize}
    \item \emph{Input:} We consider $n$ users who tested positive and each of them has visited at most $m$ places. We assume that each traced place object contains an opaque ID that refers to either a static place, e.g. a subway station, or to a dynamic place, e.g., a subway car. The identifiers are consistent across the platform so that all users have the same opaque ID for the same place.
    \item \emph{Goal:} The MA wants to have daily access to the aggregate result of contaminated locations, which is the counting of related traced place objects. From previous findings in manual tracing results~\cite{fishstore} it has been observed that it is typical to have several locations where a lot of carriers have passed by. The goal is to summarize this information concisely through a heat-map that describes the risk of each location.  
\end{itemize}

In order to securely find high-risk places, we consider two \emph{privacy-preserving solutions with different trade-offs}.
The parameters that we consider are: the communication and time complexity of the protocol, the scalability with respect to the number of users, and the security guarantees. 
Depending on the system specifications, the platform can be adjusted to the appropriate design. 
The two solutions are:

\textbf{Heavy Hitters Based Approach.} If we consider high-risk places as the frequent items from an open set, the setup resembles the heavy hitters problem. In particular, one is given a list of $m$ integers in the domain $[1, \ldots, \ell]$, and the goal is to identify the items among the domain that appear frequently in the list. In \sysname, there are $n$ users, each user has a set of visit record objects/items. If we consider the MA as the untrusted aggregator, the MA would like to continuously track the most recent heavy hitters (i.e. the popular items/places), namely those items that were held by at least a threshold number of different users. 
The problem of computing heavy hitters has been well-studied in the area of data stream algorithms. The earliest work can be traced back to the 1980s~\cite{BoyerHeavyhitter81}. Over the last few years, several works propose  efficient privacy-preserving schemes that can achieve practical performance~\cite{BassilyNeurIPs17}. However, most of the solutions rely on  differentially private guarantees. 

\textbf{Oblivious Aggregation Approach.} We are also interested in a secure computation model where no information about the private data held by the parties can be inferred during the execution of the protocol.  Given an array of key-value pairs, oblivious aggregation allows to compute some aggregation function over all pairs with the same key in the privacy-preserving manner. 
We assume that there exists a one-to-one index mapping table of size $\ell$ between a particular place and the index (e.g. zip code area). Each diagnosed carrier has a set of (key, value) where the key is an index $i \in [1, \ldots, \ell]$ which has been mapped to a visit traced place in the mapping table, and the value indicates how many time the carrier visited this place. Given $n>2$ diagnosed carriers (users), each holding a private set of key-value pairs, the functionality of oblivious aggregation is to allow the MA to compute the sum of all values with the same key without revealing any information. The oblivious aggregation~\cite{Boyle2016, SOda19} can be implemented in time $O(\ell \log \ell)$ per each user's query.

\section{Conclusion}

In this paper, we propose \sysname, a privacy-preserving contact tracing platform that breaks data silos. We show that by including business-side participation and standardizing the format of data, we can incorporate existing efforts and improve coverage. By deploying the state-of-the-art cryptographic protocols from the area of secure multi-party computation, we can achieve an efficient design that meets the privacy needs of contact tracing. We motivate our design choices by a complete workflow of a multi-source decentralized approach.

Our vision for a privacy-preserving data collaboration platform can be extended \emph{beyond the problem for contact tracing}. 
The design principles of our platform can be adapted to address other pressing problems such as private contact discovery~\cite{Kales2019} as well as compromised credential checking~\cite{Li2019}. 
The contact tracing scenario described in this work can be the first step towards a practical decentralized secure data collaboration platform.

\section*{Acknowledgments}
We thank Peng Gao, Kareem Shehata, Jiaheng Zhang, Tiancheng Xie, and Xinyun Chen for helpful discussion. This material is in part based upon work supported by the National Science Foundation(NSF) under Grant No. TWC-1518899, DARPA under Grant No. N66001-15-C-4066, Center for Long-Term Cybersecurity (CLTC), and IC3 industry partners. Any opinions, findings, and conclusions or recommendations expressed in this material are those of the author(s) and do not necessarily reflect the views of NSF, DARPA, CLTC or IC3.

\end{document}